\documentclass[a4paper,11pt]{article}

\pdfoutput=1 
\usepackage{jcappub} 
\usepackage{eurosym}

\usepackage[T1]{fontenc} 

\usepackage{amssymb}
\usepackage[dvipsnames]{xcolor}
\usepackage[normalem]{ulem}
\usepackage{systeme}
\usepackage{bm}
\usepackage{verbatim}
\usepackage{amsmath}
\usepackage[makeroom]{cancel}
\usepackage{hyperref}
\newcommand{\be}{\begin{equation}}
\newcommand{\ee}{\end{equation}}
\newcommand{\bea}{\begin{eqnarray}}
\newcommand{\eea}{\end{eqnarray}}

\def\d{{\rm d}}
\def\H{{\mathcal H}}

\def\k{\boldsymbol{k}}

\def\odm{{\Omega_{\rm cdm  0}}}
\def\ob{{\Omega_{\rm b 0}}}

\def\fnl{f_{\rm NL}}

\title{Multi-wavelength spectroscopic probes: biases from neglecting light-cone effects}

\author[1]{Jan-Albert Viljoen,}
\author[2,1]{Jos\'e Fonseca,}
\author[1,3,4]{Roy Maartens}
\affiliation[1]{Department of Physics \& Astronomy, University of the Western Cape,\\Cape Town 7535, South Africa}
\affiliation[2]{School of Physics \& Astronomy, Queen Mary University of London, London E1 4NS, UK}
\affiliation[3]{Institute of Cosmology \& Gravitation, University of Portsmouth,  Portsmouth PO1 3FX, UK}
\affiliation[4]{National Institute for Theoretical and Computational Sciences (NITheCS), South Africa}

\emailAdd{javiljoen74@gmail.com}

\abstract{Next-generation cosmological surveys will observe larger cosmic volumes than ever before, enabling us to access information on the primordial Universe, as well as on relativistic effects. In a companion paper, we applied a Fisher analysis to forecast the expected precision on $\fnl$ and the detectability of the lensing magnification and  Doppler contributions to the power spectrum. Here we assess the bias on the best-fit values of $\fnl$ and other parameters, from neglecting these light-cone effects. We consider forthcoming 21cm intensity mapping surveys (SKAO) and optical galaxy surveys (DESI and Euclid), both individually and combined together. We conclude that lensing magnification at higher redshifts must be included in the modelling of spectroscopic surveys. If lensing is neglected in the analysis, this produces a bias of more than 1$\sigma$~-- not only on $\fnl$, but also on  the standard cosmological parameters.}

\begin{document}
\maketitle
\flushbottom

\section{Introduction}\label{sec:intro}

An important scientific goal of forthcoming galaxy surveys is to tighten constraints on primordial non-Gaussianities with large-scale structure. In a companion paper \cite{Viljoen:2021ypp}, we used a Fisher forecast to estimate the constraints on the local-type primordial non-Gaussianity parameter $\fnl$, from its effect on the clustering bias of future surveys with high redshift resolution:
a bright galaxy sample (similar to the BGS survey with the Dark Energy Spectroscopic Instrument (DESI)   \cite{2016arXiv161100036D}), an H$\alpha$ survey (similar to that of Euclid \cite{Euclid:2019clj}), and 21cm intensity mapping surveys like those planned for the SKAO-MID telescope, in lower- and higher-frequency bands, denoted IM1 and IM2  \cite{Bacon:2018dui}. This multi-wavelength choice of four surveys is motivated by: high redshift resolution in order to detect the Doppler effect; good coverage of redshifts in the range $0<z<2$; a negligible cross-shot noise between optical and 21cm intensity samples; and the very different systematics affecting optical and 21cm radio surveys, which are suppressed in cross-correlations.
We showed in \cite{Viljoen:2021ypp} that the full combination of four surveys could achieve a constraint of $\sigma(\fnl)\sim 1.5$, detect the Doppler effect with a signal-to-noise ratio $\sim$8, and isolate the contribution from lensing magnification with $\sim$2\% precision. {Basic properties of the surveys and the uncertainties on $\fnl$ and relativistic effects are shown in \autoref{tab1}.}

\begin{table}[!h]
\caption{{Survey volumes and uncertainties on $\fnl$ and relativistic  terms in \eqref{dobs}, marginalising over the standard cosmological parameters (excluding priors). $\sigma(\varepsilon_{\rm P})$ is always $\gg 1$ and is not shown.  Uncertainties when marginalising {\em also}  over the Gaussian clustering biases in each bin, $b_A(z_i)$, are in brackets. $\otimes$ is a multi-tracer combination and $\oplus$ is a sum of Fisher information. (From  \cite{Viljoen:2021ypp}.) }}\label{tab1}
\centerline
{
\begin{tabular}{clccccc}
\\ \hline\hline
Redshift  & Survey & $\Omega_{\rm sky}/{\rm deg}^2$ & {$\sigma(\fnl)$} &  {$\sigma(\varepsilon_{\rm D})$}  & {$\sigma(\varepsilon_{\rm L})$} \\
\hline
$0.1-0.58$ & {BGS (DESI-like)} & 15,000 & 26.36 (26.38)  & 7.57 (7.57)  & 0.32 (0.39)  \\
& {IM2 (SKAO-like)}  & 20,000 & 35.33 (35.74) & 18.04 (18.07) & $-$   \\
& {IM2$\otimes$BGS}  &10,000  & 2.10 (2.12) & 0.14 (0.14)  & 0.12 (0.13)  \\
\hline
$0.9-1.8$ & {H$\alpha$ (Euclid-like)}  &15,000 & 9.32 (9.34)  & 9.08 (9.08) & 0.04 (0.04)  \\
$0.35-3.05$ & {IM1 (SKAO-like)} & 20,000 & 4.65 (4.72) & 6.28 (6.29) & $-$   \\
$0.60-3.05$ & {IM1$\otimes$H$\alpha$} & 10,000 & 3.05 (3.06) & 0.37 (0.37) & 0.03 (0.03)  \\
\hline
$0.1-3.05$ & {(IM2$\otimes$BGS)$\oplus$(IM1$\otimes$H$\alpha$)} & 10,000 & 1.70 (1.70)  & 0.13 (0.13)  & 0.03 (0.03) \\
& {IM2$\otimes$BGS$\otimes$IM1$\otimes$H$\alpha$} & 10,000 & 1.55 (1.55)  & 0.13 (0.13)  & 0.02 (0.02) \\
\hline\hline
\end{tabular}
}
\end{table}

{\autoref{tab1} shows that marginalisation over the clustering biases degrades the single-tracer constraints slightly, but has a negligible effect on multi-tracer constraints. This follows from the very large number of redshift bins and consequent cross-bin correlations.} 

Relativistic light-cone effects are usually neglected in the modelling of galaxy and 21cm intensity mapping surveys. This has been mainly a reasonable approximation up to now, but next-generation surveys will make higher demands on theoretical accuracy. In previous galaxy surveys with smaller survey sizes, or at lower redshifts, or with shot-noise-dominated samples, it was safe not to consider the light-cone corrections in the number density contrast. But as future surveys probe larger volumes, go to higher redshifts, and reduce shot-noise, it is important to understand whether neglecting such effects in the modelling can bias our measurements of parameters.

The observed number density/ temperature contrast $\Delta_A$ is given by the contrast at the source $\delta_{A}$, modulated not only by redshift-space distortions, but also by lensing magnification and other relativistic effects (see  \cite{Viljoen:2021ypp} for more details):
\bea \label{dobs}
 \Delta_A  
=\delta_A + \Delta^{\rm RSD} + \varepsilon_{\rm D}\Delta^{\rm Dopp}  + \varepsilon_{\rm L}\Delta^{\rm Lens}  + \varepsilon_{\rm P}\Delta^{\rm Pot}\,.
\eea
The $\varepsilon_{\rm I}$ parameters have fiducial value 1, corresponding to the correct theoretical expression. The errors $\sigma(\varepsilon_{\rm I})$ give an estimate of the detectability of these contributions. The Doppler term,  
$\Delta^{\rm Dopp}={[b_{\rm e}-5s+(5s-2)/\H\chi+\d\ln\H/\d\ln(1+z)]}\bm v\cdot \bm n$, is sourced by radial peculiar velocities; 
the lensing term, $\Delta^{\rm Lens}={(5s-2)}\kappa$, is sourced by the lensing convergence, and the potential term is sourced by Sachs-Wolfe, integrated Sachs-Wolfe, and time-delay effects. {Here $s$ is the magnification bias (with $5s-2=0$ for intensity mapping) and $b_{\rm e}$ is the evolution bias.}

There is an important distinction between the lensing term and the other two relativistic terms:
\begin{itemize}
\item
{The lensing convergence
 $\kappa$ receives contributions from matter fluctuations $\delta_{\rm m}$ all along the  line-of-sight  and therefore mixes different scales \cite{Villa:2017yfg}. 
This means that the scale dependence of $\kappa$ differs from that of $\delta_{\rm m}$. As a result, the lensing term $\Delta^{\rm Lens}$ is not fully degenerate, but  partially degenerate with the  clustering term $\delta_{A}=b_A\,\delta_{\rm m}$ \cite{Namikawa:2011yr,Lorenz:2017iez,Jelic-Cizmek:2020pkh}.}
\item
The Doppler term scales in Fourier space as $\Delta^{\rm Dopp} \sim \delta_{\rm m}\,H_0/k$. It is only non-negligible on ultra-large scales.
\item
The potential term scales in Fourier space as $\Delta^{\rm Pot} \sim \Phi\sim \delta_{\rm m}\,H_0^2/k^2$, where $\Phi$ is the Bardeen potential. This is only non-negligible on even larger scales than the Doppler term.
\item
{In correlations, the leading Doppler contribution is  $\delta_{\rm m}\times\mbox{Doppler} \sim(\delta_{\rm m})^2\,H_0/k$, which dominates over the leading potential contribution,  $\delta_{\rm m}\times\Phi\sim (\delta_{\rm m})^2\,H_0^2/k^2$.}
\item
The ultra-large scale relativistic effects are partially degenerate with the contribution of  scale-dependent clustering bias from $\fnl$. A simple model of scale-dependent bias is given by (see \cite{Barreira:2020ekm} for improved models):
\be\label{eq:bng}
\delta_{\rm g}(z,\k)=\left[b(z)+3 {\fnl}\, \frac{ \delta_{\rm crit}\big[b(z)-1 \big]{D(z_{\rm d})(1+z_{\rm d})\,\Omega_{\rm m0} H_0^2} }{D(z)\, T(k) \, k^2}\right]\delta_{\rm m}(z,\k)\,,
\ee
where  $\delta_{\rm crit}=1.686$ is the threshold density contrast for spherical collapse, $T$ is the matter transfer function (normalised to 1 on ultra-large scales),  $D$ is the growth factor (normalised to 1 at $z=0$), and $z_{\rm d}$ is the redshift at decoupling. 
\end{itemize}

{The Doppler term dominates over the potential contributions on ultra-large scales. Since it has $H_0/k$ and $(H_0/k)^2$ contributions, it is partially degenerate with the effect of $\fnl$, and it follows that neglecting the Doppler contribution could bias future measurements of $\fnl$, as pointed out in \cite{Bruni:2011ta,Jeong:2011as,Bertacca:2012tp,Camera:2014sba,Lorenz:2017iez,Contreras:2019bxy,Bernal:2020pwq,Martinelli:2021ahc} (see also \cite{Kehagias:2015tda,DiDio:2016gpd,Koyama:2018ttg,Maartens:2020jzf} for the galaxy bispectrum). The degree of bias is determined by the amplitudes of the Doppler and $\fnl$ contributions, which depend on the surveys considered.} Neglecting the lensing contribution can also lead to a biased measurement of $\fnl$ -- by biasing the estimate of clustering bias that affects the amplitude of the $\fnl$ contribution \cite{Namikawa:2011yr,Lorenz:2017iez}. In general, the conclusion is that one needs to include the relativistic effects in the modeling to avoid a biased estimation of the $\fnl$ best-fit value. However, the degree of bias depends on the survey specifications. One should note that $\fnl$ is also sensitive to bias from observational systematics such as stellar contamination \cite{Rezaie:2021voi} and foreground contamination of 21cm intensity maps \cite{Cunnington:2020wdu,Fonseca:2020lmi}. The analysis in our companion paper \cite{Viljoen:2021ypp} takes into account these systematics, and shows that their effect on $\sigma(\fnl)$ is considerably reduced in multi-tracer constraints.

In addition to primordial non-Gaussianity, we can ask whether neglecting the relativistic effects will also bias the measurements of standard cosmological parameters \cite{Montanari:2015rga,Cardona:2016qxn,Villa:2017yfg,Lorenz:2017iez,Camera:2018jys,Jelic-Cizmek:2020pkh}. The overall conclusion of previous work is that the neglect of lensing can produce a significant bias, while the neglect of Doppler and potential contributions has negligible impact. Our results are consistent with this.

To fully include all relativistic effects, all radial  correlations of redshift bins, and all wide-angle effects, we use  the angular power spectra
\be
C^{AB}_\ell ( z^A_i,z^B_j )= 4\pi\int{\rm d} \ln k\, \Delta_{A\ell}( z^A_i,k)\, \Delta_{B\ell}( z^B_j,k)\, \mathcal P_\zeta (k)\,, \label{eq:Cl}
\ee
which also do not require any Alcock-Paczynski correction.
Here $\mathcal P_\zeta (k)$ is the primordial power spectrum, and $A,B$ denote tracers.  We study the importance of including relativistic effects in single- and multi-tracer cases, and at low and high redshifts. It turns out that for the surveys we consider, lensing magnification (at higher redshifts) must be included for $\fnl$ and standard cosmological parameters. Omitting the lensing contribution produces a signficant bias ($>1\sigma$)  on the best-fit value of $\fnl$ and  standard cosmological parameters. The bias induced by neglecting the Doppler effect is below $1\sigma$ for all parameters, but since this effect has signal-to-noise $\sim8$, it should in principle be included.

In \autoref{sec:biasTH} we review how to estimate the bias on best-fit values of parameters. Results are given in \autoref{sec:results} and we conclude in \autoref{sec:conclusion}. Our fiducial cosmology is given by: $A_{\rm s}=2.142\times 10^{-9}$, $n_{\rm s}=0.967$, $\odm=0.26$, $\ob=0.05$, $w=-1$, $H_0=67.74\,$ km/s/Mpc, $\fnl=0$, where $\Omega_{\rm cdm0}+\Omega_{\rm b0}=\Omega_{\rm m0}$.
{We use redshift bins of width $\Delta z=0.03$; a minimum multipole  $\ell_{\rm min}=5$ to avoid large-scale systematics in the galaxy surveys (e.g. stellar contamination) and the intensity surveys (foreground contamination); and a maximum multipole $\ell_{\rm max}(z)=0.2h(1+z)^{2/(2+n_s)}\chi(z)$, to stay in the linear-perturbation regime. }

\section{Estimating the bias on parameter measurements} \label{sec:biasTH}

One can estimate biases on the best-fit values of parameters from unaccounted systematics or incomplete theoretical modelling in broadly two ways. In one approach, one can use the `correct' mock data and do inference on this data using the `wrong' model (see \citep{Martinelli:2021ahc} for a recent example). This approach provides a precise evaluation of the biasing introduced but is time-consuming and computationally expensive. We follow the alternative approach, which uses the Fisher matrix formalism in the case of nested model selection \cite{Heavens:2007ka}. 

In the Fisher formalism we assume that the posterior distribution of a set of $m$ parameters $\bm{\vartheta}$, given some $n$ dimensional data vector $\bm{X}$, is Gaussian: 
\be
 \mathcal{P} ({\bm\vartheta} | {\bm X} )= {\left[(2\pi)^m \det {(\bm F^{-1})}\right]^{-1/2}}\, \exp\left[-\frac12 ({\bm \vartheta}-\bar {\bm \vartheta})^{T}{\bm F}\,({\bm \vartheta}-\bar {\bm \vartheta})\right]\,.
\ee
The best-fit values $\bar{\bm \vartheta}$ maximise the posterior. The Fisher matrix $\bm{F}$ is the inverse of the covariance of the parameters and the marginal error of a parameter is given by
\be
    \sigma^2(\vartheta_\alpha)=\left[{\bm F}^{-1}\right]_{\vartheta_\alpha\vartheta_\alpha}\,.
\ee
 Using Bayes theorem, we relate the posterior with the likelihood of the data ${\cal L}$ via
\be
{\cal P} ({\bm\vartheta} | {\bm X} )\propto {\cal L} ({\bm X} | \bm\vartheta ) \, {\Pi} (\bm\vartheta )\,,
\ee
where $\Pi$ is the prior. For simplicity we do not consider any priors on the parameters. Assuming the likelihood to be Gaussian,
\be
{\cal L} ({\bm X} | \bm\vartheta )= \left[(2\pi)^n \det {\bm \Gamma}\right]^{-1/2}\,\exp\left[-\frac12 ({\bm X}-\bar {\bm X})^{T}{\bm\Gamma}^{-1}({\bm X}-\bar {\bm X})\right]\,,
\ee
where $\bm\Gamma$ is the covariance of the data, and $\bar{\bm X}$ is the value that maximises the likelihood.  
Since both $\bar {\bm X}$ and $\bm\Gamma$ can depend on $\bm\vartheta$, one sees that the modelling between the data and the parameters we want to fit influences our estimates.

Making incorrect {model} assumptions will {therefore} lead to a shift in the {best-fit value} of the parameters considered. In the example of nested models, there is a sub-model with parameters $\psi_i$, so that $\vartheta_\alpha=\{ \psi_i, \varphi_I\}$. Suppose that we make an incorrect assumption, and fix the values of $\varphi_I$ to  incorrect values $\hat\varphi_I$, instead of the true values $\bar{\varphi}_I$. This shifts $\varphi_I$ by the amount,
\be\label{eq:fixed_bias}
\delta \varphi_I = \hat{\varphi}_I - \bar{\varphi}_I \,.
\ee
As a consequence, we will estimate the remaining parameter  values to be $\hat{\psi}_i$, instead of their true  values $\bar{\psi}_i$. The bias on the best-fit values is:
\be\label{eq:best_fit}
\delta \psi_i = \hat{\psi}_i -\bar{\psi}_i \,.
\ee
We relate \eqref{eq:best_fit} to \eqref{eq:fixed_bias} through the covariance of the parameters, determined by the amount of information we expect to extract from the survey \cite{Camera:2014dia,Fonseca:2020lmi}:
\be\label{eq:best_fit_def}
\delta \psi_i = -\sum_{I,j}\,\delta\varphi_I\, F_{\varphi_I \psi_j} \,\left[ {\bm H}^{-1}\right]_{\psi_j \psi_i} \, ,
\ee
where $\bm{F}$ is the Fisher matrix of the total set $\vartheta_\alpha$, and $\bm{H}$ is the Fisher matrix of the sub-model with the parameters $\psi_i$ that we wish to fit. 

The angular power spectra are given by the covariance of the maps ${\bm X}=[a_{\ell m}^{A}(z^A_i)]$, where $A$ ranges over the different tracers and $i$ over the redshift bins. The Fisher matrix is (see  \cite{Viljoen:2021ypp}): 
\bea \label{eq:fishercl}
F_{{\vartheta}_\alpha{\vartheta}_\beta}=\sum_{\ell_{\rm min}}^{\ell_{\rm max}} \frac{(2\ell+1)}{2 }f_{\rm sky}\,{\rm Tr}\Big[ \big(\partial_{{\vartheta}_\alpha} \bm{C}_{\ell}\big)\, \bm{\Gamma}_\ell^{-1} \big(\partial_{{\vartheta}_\beta}\bm{C}_{\ell}\big)\,\bm{\Gamma}_\ell^{-1}\Big]\,,
\eea
where
\be\label{eq:matrix_Cl_Gamma}
\bm{\Gamma}_\ell 
%=\big[ \Gamma_\ell(z_i,z_j)\big]
=\bm{C}_\ell+\bm{\mathcal N}_\ell\,.
\ee
The noise $\bm{\mathcal N}$ depends on the survey: shot-noise in the case of galaxy surveys,  instrumental noise in the case of intensity mapping. We also need to incorporate the effects of the telescope beam on intensity mapping. (More details are given in  \cite{Viljoen:2021ypp}).

For the forecasts we use the parameters
\be \label{pars}
{\vartheta}_{\alpha}=\Big\{ \fnl,\, \varepsilon_{\rm D},\, \varepsilon_{\rm L},\, \varepsilon_{\rm P};\, \Omega_{\rm m0},\,  w,\, n_{\rm s},\, H_0,\, A_{\rm s};\, {b_A(z_i)} \Big\}\,,
\ee
{where $b_A(z_i)$ are the Gaussian clustering bias values for each tracer in  each redshift bin.}
We are interested in the effect of neglecting the Doppler, lensing and  potential effects, so that 
\bea
\varphi_I=\big\{\varepsilon_{\rm D},\, \varepsilon_{\rm L},\, \varepsilon_{\rm P}\big\}\,.
\eea
Neglecting the relativistic effects means setting $\hat\varepsilon_{\rm I}=0$, which means a shift from the true values of  
\be
\delta \varepsilon_{\rm D}=\delta \varepsilon_{\rm L}=\delta \varepsilon_{\rm P}=-1\,.
\ee
We then determine the bias on the best-fit values of the remaining parameters:
\be
{\psi}_i= \big\{ \fnl;\,  \Omega_{\rm m0},\,  w,\, n_{\rm s},\, H_0,\, A_{\rm s} \big\}\,.
\ee
This bias is best expressed as normalised by the errors $\sigma(\psi_i)$. We thus define the normalised biases from neglecting individual relativistic effect as
\be \label{eq:deltaGR_def}
\delta^{I}\psi_i\equiv\frac{\delta\psi_i(\varphi_I)}{\sigma(\psi_i)}\,,
\ee
where $\delta\psi_i(\varphi_I)$ denotes the case of \eqref{eq:best_fit_def} when $I$ is fixed at one value and only $j$ is summed over. It follows that the normalised bias from all relativistic effects combined is
\be \label{delrel}
\delta^{\rm rel}\psi_i=\frac{\delta\psi_i}{\sigma(\psi_i)}\quad \mbox{where}\quad \delta\psi_i=\sum_I\delta\psi_i(\varphi_I)\,.
\ee

In general, if $|\delta^{I}\psi_i|<1$, then any induced bias is smaller than the error bars and can be safely neglected. On the other hand, if $|\delta^{I}\psi_i|>1$ then the effect $I$ should not be neglected in the model. Note however that the approximation made in deriving \eqref{eq:best_fit_def} breaks down when  $|\delta^{I}\psi_i|> 1$.  For biases above $1\sigma$, the value of $\delta^{I}\psi_i$ is not reliable~-- it does not quantify the bias, but qualitatively it confirms that the bias is larger than the error bars. This is sufficient for our purposes.

\section{Results} \label{sec:results}

\begin{table}%[!ht]
\caption{Marginal errors (in percentage) on standard cosmological parameters, for individual surveys and in multi-tracer ($\otimes$) combination, including all relativistic effects. $\oplus$ denotes the sum of independent multi-tracer pairs (excluding the IM2 band overlap in band IM1). Results exclude priors.}
\label{tab:constraints_CP}
\centerline
{
\begin{tabular}{clccccc}
\\ \hline\hline
 Redshift & Survey  & $\Omega_{\rm {m0}}$ & $n_{\rm s}$ & $H_0$  & $w$ &  $A_{\rm s}$ \\
\hline
$0.10-0.58$ &{BGS {(DESI-like)}} & 2.34  & 5.95  & 12.43  & 5.25  & 19.41\\
& {IM2 {(SKAO-like)}} & 3.57  & 8.89  & 18.57  & 4.85  & 28.03\\
& {IM2$\otimes$BGS}  & 1.14  & 2.04  & 3.78  & 3.30  & 5.96\\
\hline
$0.90 -1.80$ & {H$\alpha$ (Euclid-like)} & 1.18  & 3.57  & 6.79  & 3.19  & 10.37\\ 
$0.35-3.05$ &  {IM1 {(SKAO-like)}} & 2.27  & 5.64  & 11.23  & 2.53  & 16.57 \\
$0.60-3.05$ &{IM1$\otimes$H$\alpha$} & 1.13  & 2.85  & 5.60  & 2.29  & 8.32\\
\hline
$0.10-3.05$ &  {(IM2$\otimes$BGS)$\oplus$(IM1$\otimes$H$\alpha$ )} & 0.69  & 1.58  & 3.08  & 1.35  & 4.61\\
& {IM2$\otimes$BGS$\otimes$IM1$\otimes$H$\alpha$} & 0.68  & 1.55  & 3.02  & 1.34  & 4.51\\
\hline\hline
\end{tabular}
}
\end{table}

\subsection{Constraints on standard cosmological parameters}

Before estimating biases on best-fit values, we determine how well the standard cosmological parameters can be constrained.  \autoref{tab:constraints_CP} presents the marginal errors as a fraction of the fiducial values, expressed in percentages. Generally the high-redshift surveys give better constraints~-- since they observe a larger volume and hence sample more scales, which reduces cosmic variance. However, in the multi-tracer combination, the low-$z$ surveys provide better constraints than the high-$z$ surveys.  The reason is that cosmic variance is effectively cancelled, thereby removing the advantage of bigger volume at high $z$. Furthermore, the low $z$ surveys have the advantage of lower noise. The exception is the dark energy equation of state $w$, which benefits from measurements before and after dark energy domination, that are available only in IM1. When all surveys are combined, the best constraints are achieved. In particular, the error on $\Omega_{m0}$ is sub-percent, while all others are a few percent. The full multi-tracer combination produces  a significant improvement in precision.

\begin{figure}
 \centering
  \includegraphics[width=\textwidth]{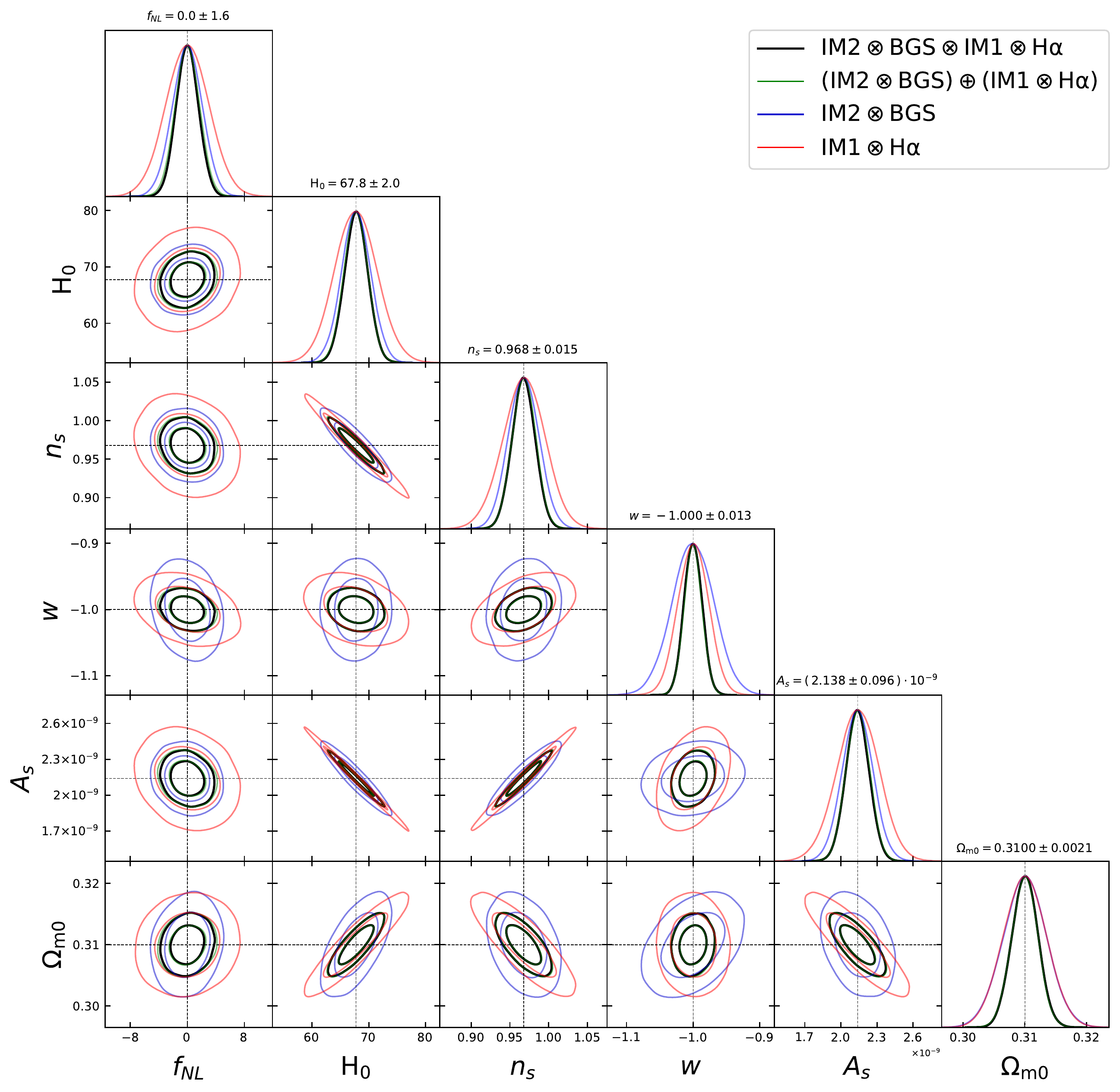}
  \caption{Contour plots of the standard cosmological parameters and $\fnl$: full multi-tracer combination of all 4 surveys  ({black});  low-$z$ multi-tracer pair (blue); high-$z$ multi-tracer pair (red); sum of low-$z$ and high $z$ Fisher information  ({green}). Fiducial values are indicated by dotted lines. (Constraints exclude priors.)} \label{fig:contour_CP}
\end{figure}

\autoref{fig:contour_CP} displays the  contour plots for the standard cosmological parameters, together with $\fnl$. Fiducial values are indicated by the dotted lines, and black contours indicate the multi-tracer correlation of all the surveys. The low- and high-$z$ multi-tracer pairs are in blue and red respectively. The sum of their Fisher information is in green. A strong degeneracy is apparent between $n_{\rm s}$, $H_0$, and $A_{\rm s}$, which is reduced as more data sets are added. By contrast, $w$  and $\Omega_{\rm m0}$, are differently degenerate with the other parameters at low and high redshifts. Except for $w$, all cosmological parameters appear to be uncorrelated with $\fnl$, which is not unexpected.

Another feature of \autoref{fig:contour_CP} and \autoref{tab:constraints_CP} is that the constraints and contours do not improve significantly when the sum of multi-tracer pairs is replaced by the full multi-tracer. This indicates that taking them as uncorrelated is a good approximation, since little information is added from low-$z$\,$\otimes$\,high-$z$ cross-correlations. The approximation considerably decreases the computation time needed.

\begin{table}%[!ht]
\caption{Bias on the best-fit value of each standard cosmological parameter (normalised by the standard deviation) that follows from neglecting relativistic effects.} \label{tab:bias_CP}
\centerline
{
\begin{tabular}{clccccc}
\\ \hline\hline
\\[-1em]
 Redshift & Survey  & $\delta^{\rm rel}\Omega_{\rm m0}$ & $\delta^{\rm rel}n_{\rm s}$ & $\delta^{\rm rel}H_0$  & $\delta^{\rm rel}w$ &  $\delta^{\rm rel}A_{\rm s}$ \\
\hline
$0.10-0.58$ & {BGS}  & 0.17  & -0.17  & 0.18  & -0.1  & -0.18\\
& {IM2}  & 0.01  & -0.01  & 0.01  & 0.0  & -0.02\\
& {IM2$\otimes$BGS}  & 0.25  & -0.30  & 0.44  & -0.67  & -0.51\\
\hline
$0.90-1.80$ & {H$\alpha$}  & 5.46  & -7.23  & 6.98  & -7.81  & -7.14\\ 
$0.35-3.05$ & {IM1} & 0.01  & -0.01  & 0.01  & -0.02  & -0.01 \\
$0.60-3.05$ & {IM1$\otimes$H$\alpha$} & 6.12  & -8.71  & 8.04  & -9.14  & -8.14\\
\hline
$0.10-3.05$ &  {(IM2$\otimes$BGS)$\oplus$(IM1$\otimes$H$\alpha$)} & 3.42  & -4.66  & 4.28  & -4.81  & -4.17\\
& {IM2$\otimes$BGS$\otimes$IM1$\otimes$H$\alpha$} & 6.07  & -7.53  & 7.25  & -7.34 & -7.05\\
\hline\hline
\end{tabular}
}
\end{table}

\subsection{Bias from neglecting relativistic effects}

We now consider the bias on the best-fit value from neglecting all relativistic effects, beginning with the standard cosmological parameters.  \autoref{tab:bias_CP} shows the biases on the best-fit of $\psi_i$, normalised to  $\sigma(\psi_i)$, that follows  from neglecting all relativistic effects in the modeling~-- i.e., $\delta^{\rm rel}\psi_i$, defined in \eqref{delrel}. At low $z$, neglecting the relativistic effects is justified, even for the multi-tracer pair IM2$\otimes$BGS.  The same is true for the high-$z$ IM1 on its own. By contrast, neglecting the relativistic effects in the H$\alpha$ survey on its own leads to significant bias for all parameters. This bias is then passed on to any multi-tracer that includes the H$\alpha$ survey. 

\begin{table}%[!ht]
\caption{Marginal error and normalised best-fit biases  $\delta^I\fnl$ on $\fnl$, from neglecting the Doppler, lensing and potential effects, and their combination, $\delta^{\rm rel}\fnl$. }
\label{tab:best_fit_bias}
\centerline
{
\begin{tabular}{clccccc}
\\ \hline\hline
\\[-1em]
 Redshift & Survey & {$\sigma(\fnl)$} &$\delta^{\rm D} \fnl$ &$\delta^{\rm L} \fnl$ &$\delta^{\rm P} \fnl$ & $\delta^{\rm rel} \fnl$ \\
\hline\hline
$0.10-0.58$ &{BGS} &26.38 & -0.05 & 0.17 & 0.02 & 0.14\\
& {IM2} & 35.74 & -0.03 & 0.0 & 0.0 & -0.03\\
& {IM2$\otimes$BGS} & 2.12 & 0.04 & 0.49 & 0.09 & 0.62\\
\hline
$0.90-1.80$ & {H$\alpha$} & 9.34 & 0.10 & 6.03 & -0.08 & 6.06\\ 
$0.35-3.05$ & {IM1} & 4.72 & 0.04 & 0.0 & -0.06 & -0.02\\
$0.60-3.05$ & {IM1$\otimes$H$\alpha$} & 3.06 & 0.08 & 3.11 & -0.15 & 3.04\\
\hline
$0.10-3.05$ & {(IM2$\otimes$BGS)$\oplus$(IM1$\otimes$H$\alpha$)} & 1.70 & 0.06 & 1.84 & -0.01 & 1.89\\
 & {IM2$\otimes$BGS$\otimes$IM1$\otimes$H$\alpha$} & 1.55 & 0.10 & 2.60 & -0.01 & 2.69 \\
\hline\hline
\end{tabular}
}
\end{table}

\begin{figure}
 \centering
  \includegraphics[width=\textwidth]{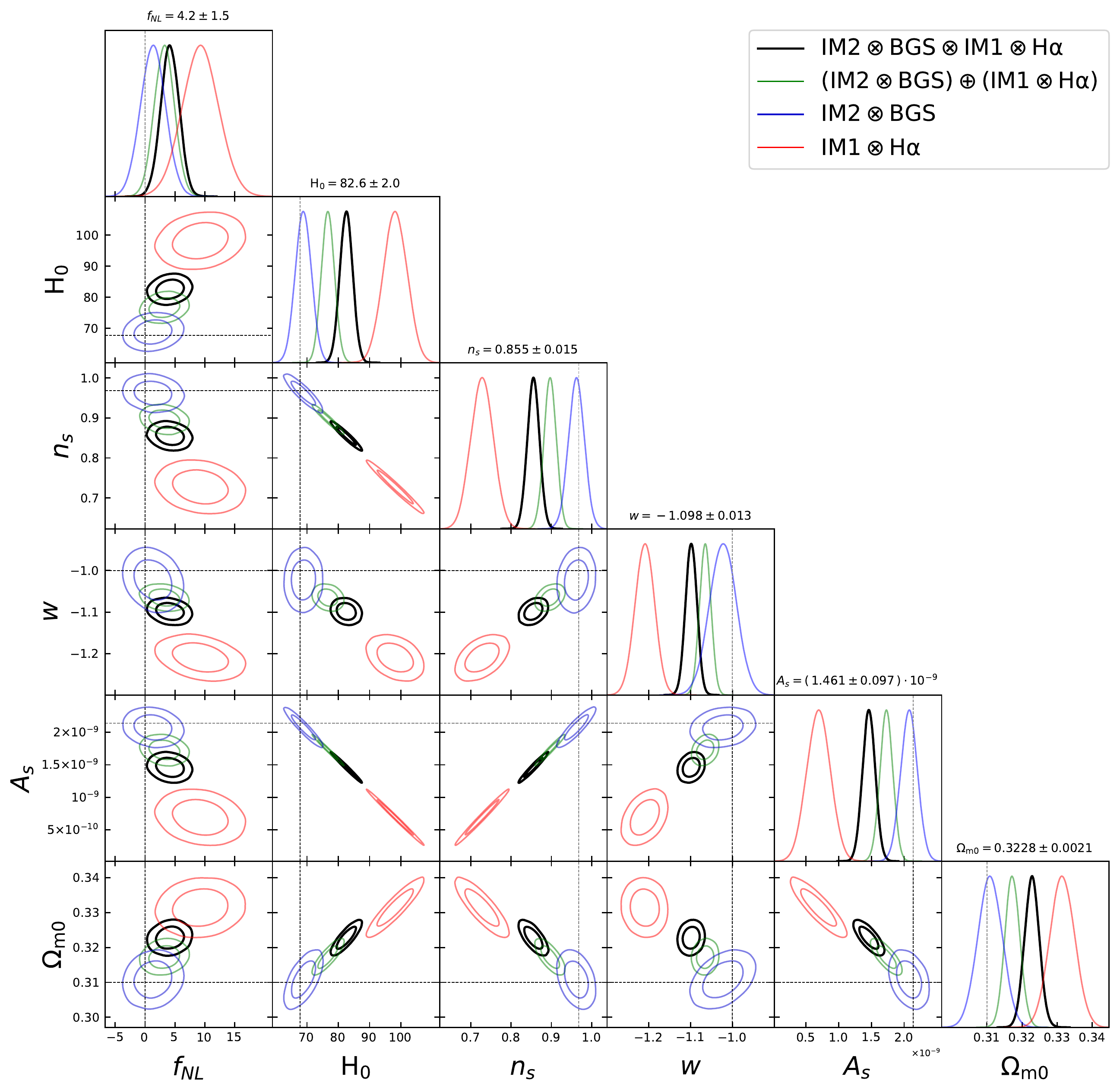}
  \caption{As in \autoref{fig:contour_CP}, but now neglecting relativistic effects, i.e., lensing, Doppler and potential effects. The best-fit values from the full multi-tracer combination (black) are given above the distributions,  showing the consequent bias on the true values (marked by dotted lines).
  }\label{fig:contour_delta_all}
\end{figure}

 \autoref{fig:contour_delta_all} shows the contours corresponding to \autoref{tab:bias_CP}. The low-$z$ multi-tracer pair and the high-$z$ multi-tracer pair disagree on the best-fit value for all parameters when relativistic effects are neglected. In other words, there is a tension between low-$z$ and  high-$z$ results, which is not eased by combining them. 
  This clearly exemplifies the problem of theoretical systematics.

In \autoref{tab:best_fit_bias}, we present the marginal error on $\fnl$ and the bias on its best-fit value, arising from neglecting the Doppler, lensing and potential effects, \eqref{eq:deltaGR_def}, and their combination, \eqref{delrel}.  The first column reproduces the results already presented in  \cite{Viljoen:2021ypp}, and serves to normalise the biases on the true value $\bar{f}_{\rm NL}=0$. The last column  is the equivalent of \autoref{tab:bias_CP} for $\fnl$. The columns in between break down the bias into the three components of the relativistic effects. Note that intensity mapping is unaffected by lensing magnification.

As in the case of the standard cosmological parameters, neglecting the relativistic effects at low redshift does not significantly bias the best-fit value of $\fnl$. Similarly, the high redshift IM1 survey does not show significant bias in $\fnl$, and it is again only the H$\alpha$ survey that suffers a significant bias on the best-fit. This $>1\sigma$ bias propagates into all multi-tracer combinations with H$\alpha$. It is apparent that the $>1\sigma$ bias is mainly due to the neglect of lensing magnification, and it follows that lensing must be included in the analysis. 

The Doppler and potential effects can lead to a bias up to $13\%$ of the error bars, in the case of IM2$\otimes$BGS. This is a significant fraction of the total 63\% bias, with lensing contributing 50\%.  If we are only interested in biases above $1\sigma$, then we can neglect the Doppler and potential contribution. However, there may be other survey combinations for which the Doppler and potential effects, when added to the lensing effect, push the bias above $1\sigma$ (or pull it below $1\sigma$).

   \begin{figure}
 \centering
  \includegraphics[width=\textwidth]{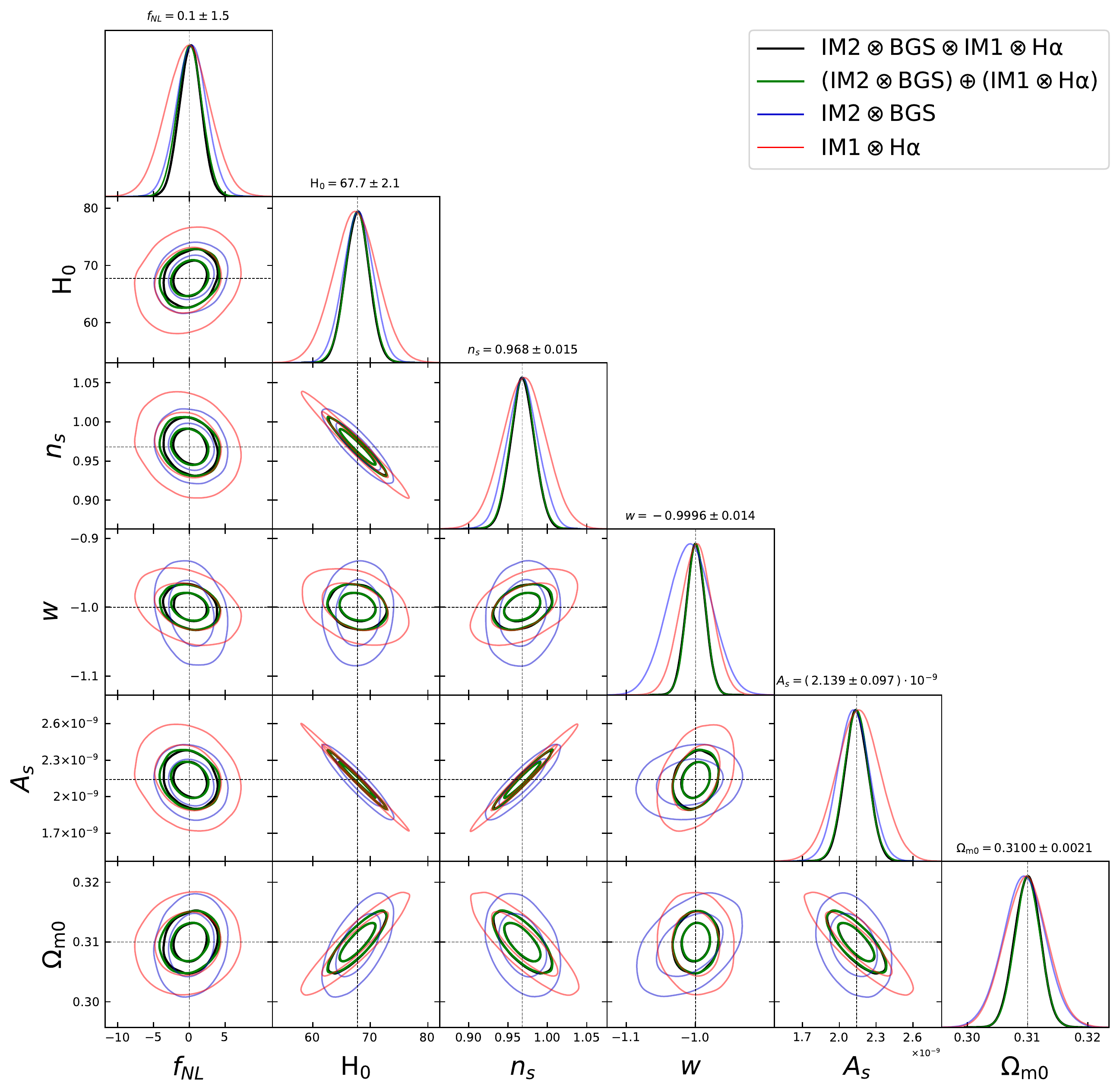}
  \caption{As in \autoref{fig:contour_delta_all}, but now neglecting the Doppler and potential effects, while including the lensing effect.
  }\label{fig:contour_delta_P}
\end{figure}

 \autoref{fig:contour_delta_all} presents the contours in the case where all relativistic effects are neglected. This incorrect model will introduce theoretical systematics in the form of a bias on the best-fit values of the parameters.  For the low-$z$ multi-tracer pair, the bias is small enough that the fiducials are still contained within the $1\sigma$ contours.   On the other hand, the high-$z$ multi-tracer pair shows strong biases in all best-fit values, in tension with the low-$z$ pair. In particular, the wrong theoretical model in the high $z$ case lead us to detect a spurious $\fnl\sim10$ at 3$\sigma$, 
  whereas the true value implies Gaussian initial conditions.  It is also apparent that when combining the low- and high-$z$ data sets, the spurious $\fnl$ detection remains ($\sim4$ at $\sim2\sigma$). 

 \autoref{tab:bias_CP}, \autoref{tab:best_fit_bias} and \autoref{fig:contour_delta_all} show that the biases in the best-fit parameters come overwhelmingly from the Euclid-like H$\alpha$ survey. It is clear that relativistic effects must be included in the modeling for theoretical accuracy.  \autoref{tab:best_fit_bias} confirms that for the surveys considered, we can safely omit the Doppler and potential effects.
 
 To confirm this, we compute the bias on the best-fit value from neglecting the Doppler and potential but keeping the lensing effect.  \autoref{fig:contour_delta_P} shows that including lensing in the modelling is sufficient to de-bias all parameters. Although some residual bias remains, it is within the 1$\sigma$ contours. Therefore we conclude that for these surveys, and their combinations, it is safe to neglect Doppler and potential effects, if the goal is to measure $\fnl$ and the standard cosmological parameters. 
 
 We emphasise that the Doppler contribution itself is detectable, with signal-to-noise of $\sim8$, and for the purpose of detection, it should be included in the modeling.  
 We do not expect any significant bias on the standard cosmological parameters from neglecting the Doppler term, since this term is only non-negligible on ultra-large scales, which contribute little to standard constraints. This is confirmed in \autoref{tab:best_fit_bias_OM} to \autoref{tab:best_fit_bias_As} below.
 {The $\fnl$ constraints do rely critically on ultra-large scales, and neglecting the Doppler effect biases the $\fnl$ best-fit by 10\% for the full multi-tracer combination of the surveys considered here (\autoref{tab:best_fit_bias}).
  There may be other combinations of surveys for which the neglect of the Doppler term  produces a more significant bias on the $\fnl$ best-fit.}

\subsection{Gaussian universe}

In a Gaussian model of the Universe, with $\fnl=0$,
do we need to include relativistic effects to avoid any bias in measurements of the standard cosmological parameters?
We summarise the results for single- and multi-tracer cases in \autoref{tab:best_fit_bias_OM} to \autoref{tab:best_fit_bias_As} for the  parameters $\Omega_{\rm m0}$,  $n_{\rm s}$, $H_0$,  $w$ and $A_{\rm s}$.

\begin{table}%[!ht]
\caption{Normalised best-fit bias in the matter density $\delta^I \Omega_{\rm m0}$, from neglecting Doppler, lensing and potential effects, and their combination $\delta^{\rm rel} \Omega_{\rm m0}$.}
\label{tab:best_fit_bias_OM}
\centerline
{
\begin{tabular}{clcccc}
\\ \hline\hline
\\[-1em]
 Redshift & Survey &$\delta^{\rm D} \Omega_{\rm m0}$ &$\delta^{\rm L} \Omega_{\rm m0}$ &$\delta^{\rm P} \Omega_{\rm m0}$ & $\delta^{\rm rel} \Omega_{\rm m0}$ \\
\hline\hline
$0.10-0.58$ & {BGS} & 0.01 & 0.17 & 0.0 & 0.18 \\
& {IM2}  & 0.02 & $-$ & 0.0 & 0.02 \\
& {IM2$\otimes$BGS}  & -0.15 & 0.4 & 0.01 & 0.26\\
\hline
$0.90-1.80$ & {H$\alpha$}  & 0.01 & 5.59 & 0.0 & 5.6\\ 
$0.35-3.05$ & {IM1}   & -0.01 & $-$ & 0.01 & 0.01\\
$0.60-3.05$ & {IM1$\otimes$H$\alpha$}  & -0.09 & 6.24 & 0.01 & 6.15 \\
\hline
$0.10-3.05$ & {(IM2$\otimes$BGS)$\oplus$(IM1$\otimes$H$\alpha$)}  & -0.01 & 3.46 & 0.0 & 3.45 \\
& {IM2$\otimes$BGS$\otimes$IM1$\otimes$H$\alpha$} & 0.0 & 6.13 & 0.0 & 6.13 \\
\hline\hline
\end{tabular}
}
\end{table}

\begin{table}%[!ht]
\caption{As in \autoref{tab:best_fit_bias_OM}, for the spectral index $n_{\rm s}$.}
\label{tab:best_fit_bias_ns}
\centerline
{
\begin{tabular}{clcccc}
\\ \hline\hline
\\[-1em]
 Redshift & Survey &$\delta^{\rm D} n_{\rm s}$ &$\delta^{\rm L} n_{\rm s}$ &$\delta^{\rm P} n_{\rm s}$ & $\delta^{\rm rel} n_{\rm s}$ \\
\hline\hline
$0.10-0.58$ & {BGS} & 0.0 & -0.18 & 0.0 & -0.19 \\
& {IM2}  & -0.01 & $-$ & 0.0 & -0.01 \\
& {IM2$\otimes$BGS}  & 0.03 & -0.32 & -0.01 & -0.30\\
\hline
$0.90-1.80$ & {H$\alpha$}   & -0.02 & -7.57 & -0.01 & -7.60\\ 
$0.35-3.05$ & {IM1}    & -0.01 & $-$ & 0.01 & 0.01\\
$0.60-3.05$ & {IM1$\otimes$H$\alpha$}  & 0.10 & -8.96 & -0.01 & -8.87 \\
\hline
$0.10-3.05$ & {(IM2$\otimes$BGS)$\oplus$(IM1$\otimes$H$\alpha$)} & 0.02 & -4.76 & 0.0 & -4.74 \\
& {IM2$\otimes$BGS$\otimes$IM1$\otimes$H$\alpha$} & 0.02 & -7.65 & -0.01 & -7.63 \\
\hline\hline
\end{tabular}
}
\end{table}

\begin{table}%[!ht]
\caption{As in \autoref{tab:best_fit_bias_OM}, for the Hubble parameter $H_0$.}
\label{tab:best_fit_bias_H0}
\centerline
{
\begin{tabular}{clcccc}
\\ \hline\hline
\\[-1em]
Redshift & Survey &$\delta^{\rm D} H_0$ &$\delta^{\rm L} H_0$ &$\delta^{\rm P} H_0$ & $\delta^{\rm rel} H_0$ \\
\hline\hline
$0.10-0.58$ &{BGS} & 0.0 & 0.19 & 0.0 & 0.20\\
& {IM2}  & 0.02 & $-$ & 0.0 & 0.02 \\
& {IM2$\otimes$BGS}  & 0.05 & 0.39 & 0.01 & 0.44\\
\hline
$0.90-1.80$ & {H$\alpha$} & 0.02 & 7.42 & 0.0 & 7.44\\ 
$0.35-3.05$ & {IM1}   & 0.0 & $-$ & 0.01 & 0.01 \\
$0.60-3.05$ & {IM1$\otimes$H$\alpha$}  & -0.10 & 8.26 & 0.01 & 8.17\\
\hline
$0.10-3.05$ & {(IM2$\otimes$BGS)$\oplus$(IM1$\otimes$H$\alpha$)} & 0.0 & 4.35 & 0.01 & 4.36 \\
& {IM2$\otimes$BGS$\otimes$IM1$\otimes$H$\alpha$} & 0.0 & 7.34 & 0.01 & 7.36 \\
\hline\hline
\end{tabular}
}
\end{table}

\begin{table}%[!ht]
\caption{As in \autoref{tab:best_fit_bias_OM},  for the dark energy equation of state $w$.}
\label{tab:best_fit_bias_w}
\centerline
{
\begin{tabular}{clcccc}
\\ \hline\hline
\\[-1em]
Redshift & Survey &$\delta^{\rm D} w$ &$\delta^{\rm L} w$ &$\delta^{\rm P} w$ & $\delta^{\rm rel} w$ \\
\hline\hline
$0.10-0.58$ &{BGS} & 0.03 & -0.14 & -0.01 & -0.12 \\
& {IM2}  & 0.0 & $-$ & 0.0 & 0.0 \\
& {IM2$\otimes$BGS}  & -0.25 & -0.43 & -0.01 & -0.69\\
\hline
$0.90-1.80$ &{H$\alpha$}  & -0.03 & -8.59 & -0.0 & -8.63\\ 
$0.35-3.05$ &{IM1}  & -0.03 & $-$ & 0.01 & -0.02 \\
$0.60-3.05$ &{IM1$\otimes$H$\alpha$}  & 0.05 & -9.63 & 0.0 & -9.58 \\
\hline
$0.10-3.05$ & {(IM2$\otimes$BGS)$\oplus$(IM1$\otimes$H$\alpha$)} & 0.07 & -4.97 & -0.02 & -4.93 \\
&  {IM2$\otimes$BGS$\otimes$IM1$\otimes$H$\alpha$} & 0.06 & -7.52 & -0.03 & -7.48 \\
\hline\hline
\end{tabular}
}
\end{table}

\begin{table}%[!ht]
\caption{As in \autoref{tab:best_fit_bias_OM},  for the  primordial power spectrum amplitude $A_{\rm s}$.}
\label{tab:best_fit_bias_As}
\centerline
{
\begin{tabular}{clcccc}
\\ \hline\hline
\\[-1em]
Redshift & Survey &$\delta^{\rm D} A_{\rm s}$ &$\delta^{\rm L} A_{\rm s}$ &$\delta^{\rm P} A_{\rm s}$ & $\delta^{\rm rel} A_{\rm s}$ \\
\hline\hline
$0.10-0.58$ &{BGS} & 0.01 & -0.21 & -0.01 & -0.21 \\
 &{IM2}  & -0.02 & $-$ & 0.0 & -0.02 \\
& {IM2$\otimes$BGS} &  -0.18 & -0.33 & 0.0 & -0.52\\
\hline
$0.90-1.80$ &{H$\alpha$}  & -0.03 & -7.71 & 0.0 & -7.74\\ 
$0.35-3.05$ &{IM1}   & 0.0 & $-$ & -0.01 & -0.01 \\
$0.60-3.05$ &{IM1$\otimes$H$\alpha$}  & 0.10 & -8.39 & -0.01 & -8.30 \\
\hline
$0.10-3.05$ & {(IM2$\otimes$BGS)$\oplus$(IM1$\otimes$H$\alpha$)} & 0.0 & -4.23 & -0.01 & -4.24 \\
& {IM2$\otimes$BGS$\otimes$IM1$\otimes$H$\alpha$} & -0.01 & -7.13 & -0.01 & -7.15 \\
\hline\hline
\end{tabular}
}
\end{table}

As expected, neglecting the Doppler and potential effects does not bias any best-fit values significantly. The largest bias is $\sim25\%$ on $w$ in the low-$z$ multi-tracer IM2$\otimes$BGS, which is below $1\sigma$. For the lensing effect, the same applies at low redshifts. Once again, it is only a Euclid-like H$\alpha$ survey, and its combinations with the other surveys, that leads to significant biases in all parameters when lensing is neglected. Although the trend is the same as found previously, the relative biases are generally smaller. This may be caused by the reduction in the parameter space volume when $\fnl$ is fixed to zero. In any case, such marginal reduction should be interpreted qualitatively, given the approximation used to compute the bias. The take-home message is that for high-$z$ spectroscopic surveys, lensing magnification must be included for unbiased measurements of the standard cosmological parameters.

\section{Conclusion} \label{sec:conclusion}

In this paper, we extended our investigation in  \cite{Viljoen:2021ypp} of the constraining power of combinations of next-generation large-scale structure surveys in the optical and radio: 
{DESI-like BGS and Euclid-like H$\alpha$ galaxy surveys, together with SKAO-like 21cm intensity mapping surveys in lower- and higher-frequency bands. Our choice was motivated by: high redshift resolution (to detect the Doppler effect); good coverage of redshifts in the range $0<z<2$; a negligible cross-shot noise between optical and 21cm intensity samples; and the very different systematics affecting optical and 21cm radio surveys.}

In \cite{Viljoen:2021ypp}, we included all relativistic observational effects on the power spectrum and used a multi-tracer analysis to forecast the precision on $\fnl$ and to determine the detectability of the relativistic effects. Here we focused on the potential theoretical systematic bias on measurements of $\fnl$ and standard cosmological parameters, which can arise if the relativistic effects are neglected in the modeling.
The observable angular power spectra  $C^{AB}_{\ell}(z_i,z_j)$ are used in the analysis, since they naturally include relativistic light-cone effects, wide-angle effects, and correlations between all redshift bins,  and do not require an Alcock-Paczynski correction. 

We first performed a Fisher analysis to estimate the expected precision on the cosmological parameters~-- using the correct theoretical model, which includes lensing, Doppler and potential effects, in addition to the standard redshift-space distortion effect.  \autoref{tab:constraints_CP} shows that the multi-tracer significantly improves on single-tracer precision, due to the combination of information and the elimination of cosmic variance. The contour plots in \autoref{fig:contour_CP} visually demonstrate the improvement in precision from combining low- and high-$z$ survey combinations, as well as showing the breaking of degeneracies between several parameters. The same qualitative features apply to the precision on $\fnl$, which was computed in \cite{Viljoen:2021ypp} and is shown in \autoref{tab:best_fit_bias}. 

Then we investigated what happens when we use the incorrect theoretical model, i.e. when we neglect one or more of the relativistic effects in the model. This leads to a theoretical systematic that threatens accuracy~-- by  biasing the best-fit (or measured) values of the parameters, as given by \eqref{eq:best_fit_def}, \eqref{eq:deltaGR_def} and \eqref{delrel}. The question is: how large is this bias for $\fnl$ and the cosmological parameters? If the bias is $<1\sigma$, the relativistic effect can be  neglected if necessary; otherwise it must be included. Each parameter in the model that is not fixed will be biased by disregarding a relativistic effect. The more free parameters there are, the greater the parameter space volume and hence the larger the potential bias. 

{The results on best-fit bias are summarised in \autoref{tab:bias_CP}--\autoref{tab:best_fit_bias_As} and \autoref{fig:contour_delta_all}.
When we separate the relativistic effects, we see that only the neglect of lensing leads to a  bias above $1\sigma$ on $\fnl$ and cosmological parameters, while the neglect of Doppler leads to at most a 25\% bias. If we are pressed to  save computation time, we can therefore neglect the Doppler and potential effects. This is confirmed in \autoref{fig:contour_delta_P}.}

\begin{figure}[!h]
 \centering
\includegraphics[width=0.9\textwidth]
    {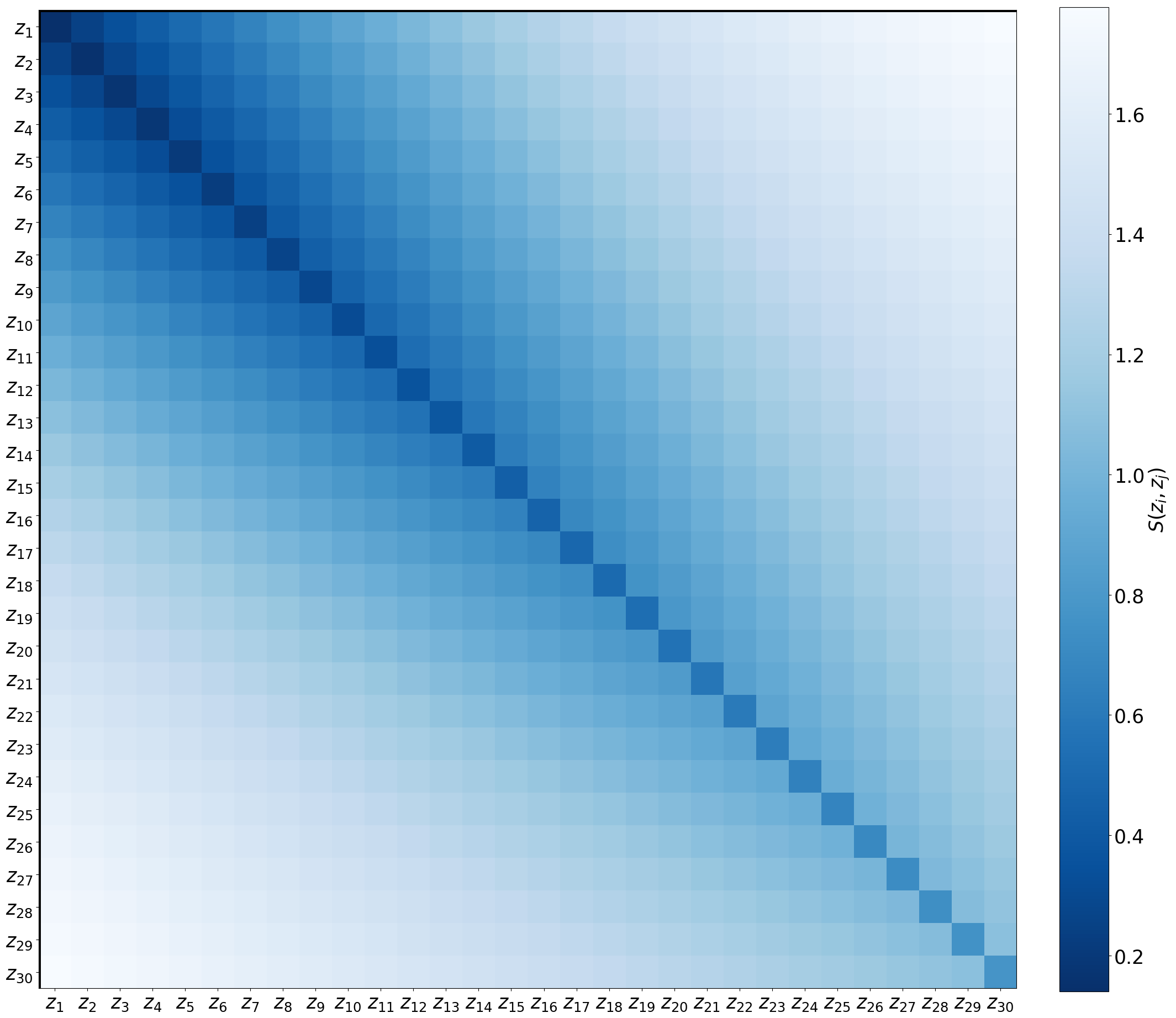}
  \caption{Signal-to-noise $S(z_i,z_j)$
  of the lensing magnification contribution in the Euclid-like H$\alpha$ survey over its redshift range $0.9<z<1.8$.} 
  \label{snrlens}
\end{figure}

{It is clear that lensing effects cannot be neglected for the full multi-tracer combination considered here -- or for any multi-tracer combination involving the H$\alpha$ survey, including the H$\alpha$ survey on its own. The special role of the H$\alpha$ survey is due  to: (a)~its high redshift reach which boosts the lensing effects, as shown in \autoref{snrlens}; (b)~the fact that the 21cm intensity surveys are unaffected by lensing magnification, although they can contribute to the lensing of galaxies in cross-correlations (see  \cite{Viljoen:2021ypp}). The BGS galaxy survey can also detect the lensing effect, but only at low significance, given its low redshift reach.} 

We confirmed that the same qualitative statements apply in the case of the bias on the cosmological parameters in a Gaussian universe, where $\fnl$ is fixed at zero. The results are summarised in \autoref{tab:best_fit_bias_OM} to \autoref{tab:best_fit_bias_As}.

One might ask how lensing drives the bias on $\fnl$, given that its signal does not require ultra-large scales in order to be significant. The point is that the lensing magnification contribution $(5s-2)\,\kappa$ is a weighted average along the line of sight of the matter density contrast~-- and therefore it
can partially mimic a change in the Gaussian clustering bias, which in turn can bias the amplitude of the $\fnl$ contribution \cite{Namikawa:2011yr,Lorenz:2017iez}.

Our results are broadly consistent with previous work on galaxy surveys, in particular
 \cite{Namikawa:2011yr,Camera:2014sba, Cardona:2016qxn,Villa:2017yfg,Lorenz:2017iez,Tanidis:2019fdh,Bernal:2020pwq}, but we consider a different combination of surveys and we use the full multi-tracer combination of four surveys. 
 
 In \cite{Jelic-Cizmek:2020pkh}, the bias on cosmological parameters is negligible for spectroscopic surveys but significant for photometric surveys. However, \cite{Jelic-Cizmek:2020pkh} uses the 2-point correlation function, without cross-bin correlations, for spectroscopic surveys. A similar result was found in \cite{Camera:2018jys} for spectroscopic surveys using the angular power spectrum in a hybrid approach which, like \cite{Jelic-Cizmek:2020pkh}, does not consider correlations from large redshift differences. By contrast, we include all cross-bin correlations amongst many thin bins. In \autoref{snrlens}, we show the lensing magnification contribution from each individual auto- and cross-bin correlation of a Euclid-like H$\alpha$ survey. The signal-to-noise ratio $S$ is given by equation (4.2) of  \cite{Viljoen:2021ypp}.  It is clear that widely separated cross-bin correlations  have the highest signal-to-noise. This accounts for our different conclusion -- and also explains why we agree with the result of \cite{Jelic-Cizmek:2020pkh} on photometric surveys, for which they do include cross-bin correlations  via a tomographic analysis.
 
 The key point is that, in order to avoid serious bias on the best-fit values of $\fnl$ and cosmological parameters, the effect of lensing magnification on the galaxy power spectrum must be included in upcoming surveys which cover high redshifts. 
 The inclusion of lensing in the theoretical modeling highlights the importance of good-precision estimates of the lensing magnification bias parameter (see also \cite{Raccanelli:2015vla,Alonso:2015uua,Montanari:2015rga,Alonso:2015sfa,Fonseca:2015laa,Cardona:2016qxn,Villa:2017yfg,Lorenz:2017iez,Jelic-Cizmek:2020pkh,Bernal:2020pwq, Maartens:2021dqy}).
 The lensing magnification and associated parameter $s$ are given by \cite{Alonso:2015uua,Maartens:2021dqy}
 \be
 \Delta^{\rm Lens}=(5s-2)\,\kappa \quad \mbox{where} \quad s={\partial \log \bar{n}_{\rm g}\over \partial m_{\rm c}} = {2\over5}\,{\phi \over \bar{n}_{\rm g}}\,.
 \ee
Here $\bar{n}_{\rm g}$ is the comoving number density at the source, which is given by an apparent magnitude integral of the luminosity function $\phi$, with limiting apparent magnitude $m_{\rm c}$. Measurements of the luminosity function, therefore, provide an estimate of $s$,  and the errors on this estimate can be modelled by simulations.
 
{The precision on the lensing  and Doppler contributions would be washed away if  $s$ and $b_e$ are poorly measured. We have partially allowed for uncertainties in $s$ and $b_e$ by marginalising over the lensing and Doppler parameters $\varepsilon_{\rm L}$ and $\varepsilon_{\rm D}$. Based on the analysis in \cite{Alonso:2015sfa}, we can estimate that errors on $s$ and $b_e$ need to be $\lesssim 10\%$ in order to preserve detectability of the lensing  and Doppler effects.}

{Finally, we note that our simplified analysis, based on Fisher forecasts, means that our  estimates of the impact of lightcone effects should be regarded as optimistic. We have  fully included   uncertainties from cosmological parameters and from the modelling of Gaussian clustering biases $b_A(z_i)$. These are important, but they have little impact on the multi-tracer, as shown in \autoref{tab1}. 
Observational systematics have not been incorporated into our analysis. Systematics on ultra-large scales include stellar contamination and dust extinction for galaxy surveys (see e.g. \cite{Rezaie:2021voi}), and foreground contamination for intensity surveys (see e.g. \cite{Alonso:2014dhk,Cunnington:2020wdu,Fonseca:2020lmi}). We have made some allowance for these systematics by excluding 
the largest scales via the cut $\ell\geq \ell_{\rm min}=5$. In Figure~7 of \cite{Viljoen:2021ypp} we showed that the full multi-tracer constraints on $\fnl$, $\varepsilon_{\rm L}$ and $\varepsilon_{\rm D}$ are robust to an increase of $\ell_{\rm min}$, up to $\sim 10-20$.}

\vfill

\acknowledgments

{We thank Ruth Durrer for a useful discussion.}
 JV and RM acknowledge support from the South African Radio Astronomy Observatory and the National Research Foundation (Grant No. 75415). RM  also acknowledges support from the UK Science \& Technology Facilities Council (STFC)  (Grant No. ST/N000550/1).
 JF acknowledges support from the UK STFC (Grant No. ST/P000592/1). This work made use of the South African Centre for High-Performance Computing, under the project Cosmology with Radio Telescopes, ASTRO-0945.

\bibliographystyle{JHEP}
\bibliography{Bibliography}

\begin{comment}
\providecommand{\href}[2]{#2}\begingroup\raggedright\endgroup
\end{comment}

\end{document}